\documentclass[fleqn]{article}
\usepackage{amsfonts}

\newcommand{\bls}[1]{\renewcommand{\baselinestretch}{#1}}

\catcode`\@=11 \@addtoreset{equation}{section}\catcode`\@=12
\def\beq#1{\begin{equation}\label{#1}}
\def\eeq{\end{equation}}

 \def\bearr{\begin{eqnarray} \lal}
 \newcommand{\bear}[1]{\begin{eqnarray}\label{#1}}
 \def\ear{\end{eqnarray}}

 \def\lal{&&\nqq {}}
 
  \def\yy{\\[5pt] {}}
 \def\nqq{\hspace*{-2em}}
 \def\dys{\displaystyle}
 \def\mst{\mathstrut}

\def\e{{\,\rm e}}

 \newcommand{\fnm}{\footnotemark}
 \newcommand{\fnt}{\footnotetext}

 \newcommand{\R}{ {\mathbb R} }

 \bls{0.957}

\begin{document}

  \begin{center}

  \large \bf On analogues of black brane solutions  in the
        model  with multicomponent anisotropic fluid
  \end{center}

 \vspace{0.3truecm}

 \begin{center}

 \normalsize\bf V. D. Ivashchuk\fnm[1]\fnt[1]{e-mail:
  ivashchuk@mail.ru},

\vspace{0.3truecm}

 \it Center for Gravitation and Fundamental Metrology,
 VNIIMS, 46 Ozyornaya ul., Moscow 119361, Russia

 \it Institute of Gravitation and Cosmology,
 Peoples' Friendship University of Russia,
 6 Miklukho-Maklaya ul., Moscow 117198, Russia

\end{center}

\begin{abstract}

 A family of spherically symmetric solutions with horizon in the
model with $m$-component anisotropic fluid  is presented. The
metrics  are defined on a manifold that contains a product of
 $n-1$ Ricci-flat ``internal'' spaces. The equation of state for
any $s$-th component is defined by a vector $U^s$ belonging to
 $\R^{n + 1}$. The solutions  are governed by   moduli
functions $H_s$ obeying non-linear differential  equations with
certain boundary conditions imposed.  A simulation of black brane
solutions in the model with antisymmetric forms is considered. An
example of solution imitating  $M_2-M_5$ configuration (in $D =11$
supergravity) corresponding to Lie algebra  $A_2$  is presented.

\end{abstract}



\section{Introduction}

In this paper we continue our investigations of
spherically-symmetric solutions with horizon (e.g., black brane
ones) defined on product manifolds containing several Ricci-flat
factor-spaces (with diverse signatures and dimensions). These
solutions appear either in models with antisymmetric forms and
scalar fields \cite{BIM}-\cite{IMtop} or in models with
(multi-component) anisotropic fluid \cite{IMS1}-\cite{DI-03}. For
black brane solutions with $1$-dimensional factor-spaces (of
Euclidean signatures) see \cite{CT,AIV,Oh} and references therein.

These  and  more general brane cosmological and spherically
symmetric solutions were obtained by reduction of the field
equations to the Lagrange equations corresponding to Toda-like
systems \cite{IMJ,IK}. An analogous reduction for models with
multicomponent anisotropic fluids was performed earlier in
\cite{IM5,GIM}. For cosmological-type models with antisymmetric
forms without scalar fields any brane  is equivalent to an
anisotropic fluid with the equations of state:

\beq{0}
  \hat{p}_i =  - {\hat \rho}  \qquad
  {\rm or} \qquad {\hat p}_i = {\hat \rho},
 \eeq

when the manifold $M_i$ belongs or does not belong to the brane
worldvolume, respectively (here ${\hat p}_i$ is the effective
pressure in  $M_i$  and ${\hat \rho}$ is the effective density).

In this paper we  present spherically-symmetric solutions with
horizon (e.g the analogues of intersecting black brane solutions)
in a model with multi-component anisotropic fluid (MCAF), when
certain relations on fluid parameters are imposed. The solutions
are governed by  a set of moduli functions $H_s$ obeying
non-linear differential master equations with certain boundary
conditions imposed. These  master equations  are equaivalent to
Toda-like equations and depend upon the non-degenerate ($m \times
 m$)  matrix  $A$.  It was conjectured earlier that the functions $H_s$
 should be polynomials when $A$ is a Cartan matrix for some semisimple
 finite-dimensional Lie algebra (of rank $m$) \cite{IMp1}.  This conjecture
 was verified  for Lie algebras: $A_m$, $C_{m+1}$, $m \geq
 1$  \cite{IMp2,IMp3}.  A special case
 of black hole solutions with MCAF corresponding to semisimple Lie
 algebra $A_1 \oplus ... \oplus A_1$  was considered earlier in
\cite{IMS2} (for  $m = 1$ see \cite{IMS1}).

 The paper is organized as follows. In Section 2 the model is
 formulated. In Section 3  spherically-symmetric MCAF solutions
 with horizon corresponding to black brane type solutions,
 are presented.  In Section 4 a polynomial structure of
 moduli functions $H_s$ for semisimple finite-dimensional
 Lie algebras is discussed.
 In Section 5 a simulation of intersecting black brane solutions is
considered and an analogue  of $M2 - M5$ dyonic solution is
presented.

\section{The model}

In this paper we deal with a family of spherically symmetric
solutions to Einstein equations with an anisotropic matter source
 \beq{1.1}
        R^M_N - \frac{1}{2}\delta^M_N R = k^2 T^M_N,
 \eeq
defined on the manifold

\beq{1.2}
\begin{array}{l}
M = {\R}_{*}\times (M_{0}=S^{d_0}) \times (M_1 = {\R})
    \times \ldots \times M_n,\\ \qquad ^{\rm radial
    \phantom{p}}_{\rm variable}\quad^{\rm spherical}_{\rm variables}
            \quad\qquad^{\rm time}
\end{array}
\eeq
with the block-diagonal metrics

\beq{1.2a}
    ds^2= \e^{2\gamma (u)} du^{2}+\sum^{n}_{i=0}
            \e^{2 \beta^i(u)} h^{[i]}_{m_i n_i }dy^{m_i}dy^{n_i }.
\eeq

Here $\R_{*} \subseteq \R$ is an open interval. The manifold $M_i$
with the metric $h^{[i]}$, $i=1,2,\ldots,n$, is a Ricci-flat space
of dimension $d_{i}$:

 \beq{1.3}
    R_{m_{i}n_{i}}[h^{[i]}]=0,
 \eeq

and $h^{[0]}$ is the standard metric on the unit sphere $S^{d_0}$,
so that

 \beq{1.4}
    R_{m_{0}n_{0}}[h^{[0]}]=(d_0-1)h^{[0]}_{m_{0}n_{0}};
 \eeq

 $u$ is a radial variable, $\kappa^2$ is the gravitational constant,
 $d_1 = 1$ and $h^{[1]} = -dt \otimes dt$.

The energy-momentum tensor is adopted in the following form for
each component of the fluid:

\beq{1.5}
 ({T^{s}}^{M}_{N})= {\rm diag} (-{\hat{\rho}^{s}},
 {\hat p}_{0}^{s}
    \delta^{m_{0}}_{k_{0}}, {\hat p}_{1}^{s}
    \delta^{m_{1}}_{k_{1}},\ldots , {\hat p}_n^{s}
    \delta^{m_{n}}_{k_{n}}),
\eeq

where $\hat{\rho}^{s}$ and $\hat p_{i}^{s}$ are the effective
density and pressures respectively, depending on the radial
variable $u$.

We assume that the following ``conservation laws''
 \beq{5.0}
 \nabla_{M}T^{(s)M}_{\ N}=0 \eeq
 are valid for all components.

We also impose the following equations of state
\beq{1.7}
    {\hat p}_i^{s}=\left(1-\frac{2U_i^{s}}{d_i}\right){\hat{\rho}^{s}},
\eeq

where $U_i^{s}$ are constants, $i= 0,1,\ldots,n$.

The physical density and pressures are related to the effective
ones (with ``hats'') by the formulae
  \beq{1.7a}
        \rho^{s} = - {\hat p}_1^{s}, \quad p_u^{s} = - \hat{\rho}^{s},
        \quad p_i^{s} = \hat{p}_i^{s} \quad (i \neq 1).
   \eeq

In what follows we put $\kappa =1$ for simplicity.

\section{Spherically symmetric solutions with horizon}

We will make the following assumptions:

\beq{2.1}
\begin{array}{l}
    1^{o}.\quad U^{s}_0 = 0 \
    \Leftrightarrow \ \hat p^{s}_0 = \hat\rho^{s} ,\yy
    2^o.\quad U^{s}_1 = 1 \
    \Leftrightarrow \  \hat p^{s}_1 = -\hat\rho^{s} ,\yy
    3^o.\quad (U^{s},U^{s})  = U^{s}_i G^{ij}U^{s}_j > 0,
    \yy
    4^o.  \quad 2(U^{s},U^{l})/ (U^{l},U^{l}) = A_{sl},
\end{array}
\eeq

 where $A = (A_{sl})$ is non-degenerate matrix,

\beq{2.2a}
    G^{ij}=\frac{\delta^{ij}}{d_i} + \frac{1}{2-D},
\eeq are components of the matrix inverse to the matrix of the
minisuperspace metric \cite{IMZ}

 \beq{2.2}
    (G_{ij}) = (d_i \delta_{ij} - d_i d_j),
 \eeq

 $i,j = 0,1,...,n$ and $D=1 + \sum\limits_{i=0}^n {d_i}$ is the
total dimension.

 The conditions $1^o$ and $2^o$ in brane terms mean that brane
``lives'' in the time manifold $M_1$ and does not ``live'' in
 $M_{0}$. Due to  assumptions $1^o$ and $2^o$ and the equations of
state (\ref{1.7}),  the energy-momentum tensor (\ref{1.5}) reads
as follows:

\beq{2.1a}
    ({T^{(s)}}^{M}_{N})= {\rm diag} (-\rho^{s},\
    \rho^{s} \delta^{m_0}_{k_0},\
    - \rho^{s}, \ p_2^{s}\delta^{m_2}_{k_2},\ \ldots ,
    p_n^{s}\delta^{m_n}_{k_n}).
\eeq

Under the conditions (\ref{1.7}) and (\ref{2.1}) we have obtained
the following black-brane-like solutions to the Hilbert-Einstein
equations (\ref{1.1}):

\bear{12} \lal
    ds^{2} = J_{0}\left( \dys\frac{\mst
    dR^{2}}{1-{2\mu}/{R^{d}}} + R^{2} d \Omega^2_{d_0} \right) -
        J_1\left(1-\frac{2\mu}{R^{d}}\right)dt^{2}
    + \sum_{i=2}^{n} J_{i} h^{[i ]}_{m_{i}n_{i}} dy^{m_{i}}dy^{n_{i}},
 \\  \lal \label{13}
    \rho^{s}= - \frac{A_s}{J_0 R^{2d_0}} \prod_{l=1}^m H_l^{-A_{sl}} ,
    \ear
 which may be derived by analogy with the black brane solutions
 \cite{IMp2,IMp3}. Here
  $d=d_0-1$,
  \beq{2.sp}
    d \Omega_{d_0}^2= h^{[0]}_{m_{0}n_{0}} dy^{m_{0}}dy^{n_{0}}
  \eeq
is the $d_0$-dimensional spherical element (corresponding to the
metric on $S^{d_0}$),

\beq{2.3}
    J_{i} = \prod_{s =1}^m H_s^{-2 h_s U^{s i }},
\eeq
 $i = 0,1,...,n$,   $\mu >0$ is integration constant and
 \bearr\label{2.4}
    U^{s i} = G^{ij}U^{s}_{j}  = \frac{U^{s}_i}{d_i} + \frac{1}{2-D}
                \sum_{j=0}^{n}U^{s}_j,
  \\ \lal \label{2.4a}
        h_s = K_s^{-1}, \quad  K_s = ({U^{s}},{U^{s}}).
\ear
 It follows from  $1^{o}$ and (\ref{2.4}) that
   \beq{2.4b}
    U^{s 0} = \frac{1}{2-D} \sum_{j=1}^{n}U^{s}_j .
  \eeq

 Functions $H_s > 0$ obey the equations

 \beq{2.2.1}
  R^{d_0} \frac{d}{dR} \left[ \left(1 -
  \frac{2\mu}{R^{d}}\right) \frac{ R^{d_0} }{H_s} \frac{d
  H_s}{dR} \right] =  B_s \prod_{l = 1}^{m}  H_{l}^{- A_{s l}},
 \eeq
 with $B_s = 2 K_s A_s$
 and the boundary conditions imposed:
  \beq{2.2.1a}
   H_s  \to H_{s0} \neq 0, \quad
    {\rm for} \quad R^{d} \to  2\mu,
   \eeq
  and
  \beq{2.2.1b}
  H_s (R = +\infty) = 1,
  \eeq
  $s = 1,..., m$.

  Here we also impose  the following (additional) condition on the solutions
   \beq{2.2.1c}
     H_s(R) > 0  \ {\rm is \ smooth \ in} \ (R_{\epsilon}, +
     \infty),
   \eeq
  $s = 1,..., m$, where $R_{\epsilon}= (2 \mu)^{1/d} e^{- \epsilon}$,
  $\epsilon > 0$.  Then  the metric (\ref{12}) has a regular horizon
  at $R^{d} =   2 \mu$ and has an asymptotically flat
  $(2 + d_0)$-dimensional section.

   Due to to (\ref{2.1}) and (\ref{2.4}) the metric reads
   \beq{12a}
    ds^{2} = J_{0}\left[ \dys\frac{\mst
    dR^{2}}{1-{2\mu}/{R^{d}}} + R^{2} d \Omega^2_{d_0}  -
       \left(\prod_{s =1}^m H_s^{-2 h_s } \right)
       \left(1-\frac{2\mu}{R^{d}}\right)dt^{2}
    + \sum_{i=2}^{n} Y_{i} h^{[i ]}_{m_{i}n_{i}} dy^{m_{i}}dy^{n_{i}} \right],
    \eeq
  where
  \beq{2.3a}
  Y_i = \prod_{s =1}^m H_s^{-2 h_s U^{s}_i/d_i }.
  \eeq

  The solution (\ref{12}), (\ref{13}) may be verified just  by
  a straightforward substitution into equations of motion.
  A detailed derivation of this solution will be given in
  a separate paper \cite{I-10}. A special
  orthogonal case when $(U^s,U^{l})= 0$, for  $s \neq l$,
  was considered earlier in \cite{IMS2} (for $m =1$ see \cite{IMS1})
  More general solutions in orthogonal case (with more general condition
  instead of $2^o$) were obtained in  \cite{DI-03}
  (for $m =1$  see \cite{DIM-02}.)

 \section{Polynomial structure of $H_s$ for  Lie algebras}

Now we deal  with solutions to second order non-linear
differential equations  (\ref{2.2.1}) that may be rewritten as
follows

\beq{5.3.1}
 \frac{d}{dz} \left( \frac{(1 - 2\mu z)}{H_s}
 \frac{d}{dz} H_s \right) = \bar B_s
 \prod_{l =1}^{m}  H_{l}^{- A_{s l}}, \eeq

 where $H_s(z) > 0$,
  $z = R^{-d} \in (0, (2\mu)^{-1})$ ($\mu > 0$) and $\bar B_s =
 B_s/ d^2 \neq 0$. Eqs. (\ref{2.2.1a}) and  (\ref{2.2.1b})
 read

 \bear{5.3.2a}
  H_{s}((2\mu)^{-1} -0) = H_{s0} \in (0, + \infty), \\
 \label{5.3.2b} H_{s}(+ 0) = 1, \ear
  $s = 1,..., m$.

 The condition  (\ref{2.2.1c}) reads as follows
   \beq{5.3.2c}
     H_s(z) > 0 \ {\rm is \ smooth \ in} \ (0, z_{\epsilon}),
   \eeq
   $s = 1,..., m$, where $z_{\epsilon}= (2 \mu)^{-1} e^{ \epsilon d}$,
   $\epsilon > 0$.

   It was conjectured in \cite{IMp1}
 that  equations (\ref{5.3.1})-(\ref{5.3.2b})
 have  polynomial solutions  when  $(A_{s s'})$ is a  Cartan matrix for
 some  semisimple finite-dimensional Lie algebra $\cal G$ of rank
 $m$.  In this case we get

 \beq{5.3.12}
 H_{s}(z) = 1 + \sum_{k = 1}^{n_s} P_s^{(k)} z^k, \eeq

 where $P_s^{(k)}$ are constants, $k = 1,\ldots, n_s$;
 $P_s^{(n_s)} \neq 0$, and

 \beq{5.2.20}
  n_s = b_s \equiv
            2 \sum_{l = 1}^m  A^{s l}
  \eeq
 $s = 1,..., m$, are the components of twice the  dual Weyl
 vector in the basis of simple  co-roots \cite{FS}.
 Here $(A^{sl}) = (A_{sl}^{-1})$.

 This conjecture  was verified for ${\bf A_m}$ and ${\bf C_{m+1}}$
 series of Lie algebras in \cite{IMp2,IMp3}. In the extremal case ($\mu
 = + 0$) an analogue of this conjecture was suggested
(implicitly) in \cite{LMMP}.

 {\bf Remark.} We note that the substitution of (\ref{1.5}),
 (\ref{1.7}), (\ref{12}),  (\ref{13}) into Hilbert-Einstein
 equations (\ref{1.1}) gives an extra equation
 \beq{5.E}
  E_{T} = \frac{d^2}{4} \sum_{s,l =1}^m h_s A_{sl}
 \left[F \frac{d}{dz} \ln H_s + \mu b_s \right]
 \left[F \frac{d}{dz} \ln H_l + \mu b_l \right]
  +  \sum_{s =1}^m A_s F \prod_{l = 1}^{m}  H_{l}^{- A_{s l}} =
  \frac{1}{2} \sum_{s =1}^m h_s b_s (\mu d)^2,
 \eeq
 where $F = 1 - 2\mu z$. $E_{T}$ is an integral of motion
 for the set of equations (\ref{5.3.1}).
 The constraint (\ref{5.E})  is satisfied identically
 due to (\ref{5.3.2a}), (\ref{5.3.2c}) (one can check this by putting
 $2\mu z = 1$).

{\bf  ${\bf A_1} \oplus \ldots \oplus {\bf A_1}$ -case.}

The simplest example occurs in the orthogonal case : $(U^s,U^{l})=
0$, for  $s \neq l$ \cite{BIM,IMJ} (see also \cite{CT,AIV,Oh} and
refs. therein). In this case $(A_{s l}) = {\rm diag}(2,\ldots,2)$
is a Cartan matrix for the semisimple Lie algebra ${\bf A_1}
\oplus \ldots \oplus {\bf A_1}$ and

 \beq{5.3.5}
 H_{s}(z) = 1 + P_s z, \eeq
 with $P_s \neq 0$,  satisfying

 \beq{5.3.5a}
 P_s(P_s + 2\mu) = -\bar B_s = - 2 K_s A_s/d^2,
 \eeq
 $s = 1,..., m$. When all $A_s < 0$ (or, equivalently, $\rho^{s} > 0$) there
 exists a unique set of numbers  $P_s > 0$ obeying (\ref{5.3.5a}).

{\bf  $A_2$-case.}

 For the Lie algebra $\cal G$ coinciding with  ${\bf A_2} = sl(3)$
  we get $n_1 = n_2 =2$ and

 \beq{5.4.1} H_{s} = 1 + P_s z + P_s^{(2)} z^{2},
 \eeq

 where $P_s=
 P_s^{(1)}$ and $P_s^{(2)} \neq 0$ are constants, $s = 1,2$.

 It was found in \cite{IMp1} that for $P_1 +P_2 + 4\mu \neq 0$
 (e.g. when all $P_s >0 $) the following relations take place

 \bear{5.4.5}
 P_s^{(2)} = \frac{ P_s P_{s +1} (P_s + 2 \mu )}{2
(P_1 +P_2 + 4\mu)}, \qquad \bar B_s = - \frac{ P_s (P_s + 2 \mu
 )(P_s + 4 \mu )}{P_1 +P_2 + 4\mu}, \ear
 $s = 1,2$.

Here we denote $s+ 1 = 2, 1$ for $s = 1,2$, respectively.

 {\bf Other solutions.}

  At the moment the ``master'' equations were integrated (using
  Maple) in    \cite{GrIvKim1,GrIvMel2} for Lie
  algebras ${\bf C_2}$ and ${\bf A_3}$, respectively.
  (For ${\bf D_4}$-polynomials in the extremal case $\mu \to +0$ see \cite{LMMP}.)

  Special solutions $H_{s}(z) = (1 + P_s z)^{b_s}$
   with $b_s$ from (\ref{5.2.20})
  appeared earlier in  \cite{Br1,IMJ2,CIM} in a context of
  so-called block-orthogonal configurations.

  {\bf Extremal case.}
  For $\mu \to +0$  the conditions
  (\ref{5.3.2a}), (\ref{5.3.2c}) should be omitted
  but we should impose the relation
   \beq{5.Ee}
    E_{T} = \frac{d^2}{4} \sum_{s,l =1}^m h_s A_{sl}
    \frac{d}{dz} \ln H_s  \frac{d}{dz} \ln H_l
    +  \sum_{s =1}^m A_s F \prod_{l = 1}^{m}  H_{l}^{- A_{s l}} =
    0,
   \eeq
   following from  (\ref{5.E}). The functions $H_s(z) > 0$
   obeying (\ref{5.3.1}) (with $\mu \to +0$) are smooth on
   $(0,+ \infty)$.    For certain relations on
   $U^s$-vectors imposed the solution under consideration has a
   horizon for  $R \to +0$ ($z \to + \infty$), e.g. it may describe
   analogues  of extremal black brane solutions \cite{IMtop}.

   When the boundary condition (\ref{5.3.2b}) is omitted we get
  a special solution with

   \beq{5.nh}
          H_s(z) = C_s z^{b_s},
   \eeq
    and
    \beq{5.C}
          C_s = \prod_{l = 1}^{m} (-b_l/\bar B_l)^{-A^{sl}},
   \eeq
   where       $b_s/\bar B_s < 0$,
       $s = 1,..., m$. The metric (\ref{12}) with
       $H_s$ from (\ref{5.nh}) has no an asymptotically flat
       $(2 + d_0)$-dimensional section.

       It should be noted that  the  solutions obeying
        $H_{s}(+ 0) = 1$ and  $b_s/\bar B_s < 0$ have an asymptotical
       behaviour $H_s(z) \sim C_s z^{b_s}$ for
       $z \to + \infty$ (e.g. in the near-horizon  limit $R \to +0$
       of  extremal black-brane-type solutions).

  \section{Examples}

 \subsection{Analogues of intersecting black brane solutions}

The solution from the previous section for MCAF allows one to
simulate the intersecting black brane solutions \cite{IMtop} in
the model with antisymmetric forms without scalar fields. In this
case the parameters $U^{s}_i$ and pressures have  the following
form:

 \beq{3.1a}
 \begin{array}{ccccccr}
 U^{s}_i & = & d_i , & p_i^{s} & = & -\rho^{s} , &  i\in I_{s};\\
    &  & 0 , &  & & \rho^{s} ,  &  i \notin I_{s}.
 \end{array}
 \eeq

Here
 $I_{s} = \{ i_1^s, \ldots,  i_{k_s}^s \} \in \{1, \ldots n \}$
is the index set \cite{IMtop} corresponding to brane submanifold
 $M_{i_1^s}  \times \ldots \times M_{i_{k_s}^s}$.

The relation  $4^o$  (\ref{2.1}) leads us to  the following
dimensions of intersections of brane submanifolds
(``worldvolumes'')  \cite{IMJ,IMtop}:

  \beq{3.1b}
  d(I_s \cap I_l)=\frac{d(I_s)d(I_l)}{D-2} +
      \frac{1}{2}K_l A_{sl}, \eeq

 $s \neq l$; $s,l = 1,\ldots ,m$.
 Here $d(I_s)$ and $d(I_l)$ are dimensions of brane worldvolumes.

  \subsection{$M_2-M_5$-analogue   for Lie algebra $A_2$}

 In \cite{IMS2} examples of MCAF-analogues of
  $M2 \cap M5$, $M2 \cap M2$, $M5 \cap M5$ black brane solutions in
  $D = 11$ supergravity, with the standard (orthogonal) intersection
  rules  were considered.

 Now we consider a solution with 2-component anisotropic
 fluid that simulates $M_2-M_5$ dyonic configuration
  in $D = 11$ supergravity \cite{IMp1}, corresponding to Lie
 algebra ${\bf A_2}$.

 The solution is  defined on the manifold
\beq{5.4.8} M =  (2\mu, +\infty
  )  \times (M_0 = S^{2})  \times (M_1 = \R) \times M_{2} \times
 M_{3}, \eeq

 where ${\dim } M_2 =  2$ and ${\dim } M_3 =  5$.
 The $U^s$-vectors corresponding to fluid components
  obey  (\ref{3.1a}) with  $I_1 = \{ 1, 2 \}$ and $I_2 = \{ 1, 3 \}$.

The solution reads as following
 \begin{eqnarray}
  g =  H_1^{1/3} H_2^{2/3} \biggl\{ \frac{dR \otimes
  dR}{1 - 2\mu / R} + R^2   h[S^2]  -
  H_1^{-1} H_2^{-1} \left(1 - \frac{2\mu}{R} \right) dt\otimes dt
   + H_1^{-1} {h}^{[2]} + H_2^{-1} {h}^{[3]} \biggr\},
    \label{5.4.9} \\
     \rho^{1}= - \frac{A_1}{J_0 R^{4}}  H_1^{-2} H_2, \qquad
   \rho^{2}= - \frac{A_2}{J_0 R^{4}}  H_1 H_2^{-2},  \label{5.4.10}
 \end{eqnarray}

 where $J_0 = H_1^{1/3} H_2^{2/3}$; $h[S^2]$ is the canonical
 metric on 2-dimensional sphere $S^2$, $h^{[2]}$ and  $h^{[3]}$ are
 Ricci-flat metrics of Euclidean signatures defined on the manifolds
 $M_2$ and $M_3$, respectively; $\mu > 0$ and $H_s$
 are defined by (\ref{5.4.1}), where $z = R^{- 1}$ and parameters
 $P_s$, $P_s^{(2)}$, $\bar B_s  = B_s = 4 A_s$ ($s =1,2$)
 obey  (\ref{5.4.5}).

This  solution simulates ${\bf A_2}$-dyon from  \cite{IMp1}
consisting of an electric $M2$-brane with a worldvolume isomorphic
to $(M_1 = \R) \times M_{2}$ and a magnetic  $M5$-brane with a
 worldvolume isomorphic to $(M_1 = \R) \times M_{3}$. The branes
are intersecting on the time manifold $M_1 = \R$. Here $K_s =
(U^s,U^s)=2$,  for all $s =1,2$.

 For the ${\bf A_2}$-dyon from \cite{IMp1} we had $\bar B_s = B_s = - 2
 Q_s^2$, where $Q_s$ is the charge density parameter of $s$-th
 brane. Thus, for fixed $Q_s$ the fluid parameters should
 obey the relations $A_s = -\frac{1}{2} Q_s^2$ and hence
 $A_s$ are negative.

  Let us consider the extremal case $\mu \to +0$  of the solution
  (\ref{5.4.9})-(\ref{5.4.10}) with $A_1 < 0$ and $A_2 < 0$.
  The near-horizon limit ($R \to +0$) gives us an exact solution
  (see (\ref{5.nh}))
 \begin{eqnarray}
  g=  C_1^{1/3} C_2^{2/3} \biggl\{ h[AdS^2] +   h[S^2]
   + C_1^{-1} {h}^{[2]} +  C_2^{-1} {h}^{[3]} \biggr\},
 \label{5.4.9a} \\
  \rho^{1}= - A_1  C_1^{- 7/3} C_2^{1/3}, \qquad
   \rho^{2}= - A_2  C_1^{2/3} C_2^{-8/3},
   \label{5.4.10a}
 \end{eqnarray}
  where $C_1 = 2  |A_1|^{2/3} |A_2|^{1/3}$ and
  $C_2 = 2  |A_2|^{2/3} |A_1|^{1/3}$,
  $h[AdS^2] =   R^{-2}(dR \otimes  dR    -  R^4 d\bar t \otimes
  d \bar t)$ is the metric of (the half of) the anti-deSitter space $AdS^2$
   (here $\bar t =  C_1^{-1/2} C_2^{-1/2} t$).
   Thus, we have obtained a static configuration defined
   on (the half of) the product  space
   $AdS^2 \times S^2 \times M_2 \times M_3$. (For the
   solution with two branes and  $C_1 = C_2 =
   1$ see \cite{I-98}.)

  {\bf Analogues of  $M2$, $M5$ and $D3$  solutions.}
   Here we outline for completeness the analogues of non-marginal
   $M2$, $M5$ ($D = 11$) and $D3$ ($D = 10$)
   black brane solutions.
   The solutions are  defined on the product manifolds
   $M =  (2\mu, +\infty )  \times (M_0 = S^{d_0})  \times (M_1 = \R) \times
   M_{2}$,  where ${\dim } M_2 = d_2 = 2, 5, 3$ and $d_0 = 7, 4, 5 $
   for $M2$, $M5$, $D3$  branes, respectively.
   The vector $U = U^1$ has the components:
    $U_0 = 0$ and $U_i = 1$ for $i > 0$. The solutions read
    as follows
   \begin{eqnarray}
   g =  H^{r}  \biggl\{ \frac{dR \otimes dR}{1 - 2\mu R^{-d}}
   +  R^2   h[S^{d_0}]  -
    H^{-1}  \left(1 - 2\mu R^{-d} \right) dt\otimes dt
   + H^{-1} {h}^{[2]}  \biggr\},
    \label{5.4.11} \\
       \rho =  \rho^{1} = - A H^{-2 - r}  R^{ - 2 d_0}  \label{5.4.12},
 \end{eqnarray}
 where ${h}^{[2]}$ is a Ricci-flat metric on $M_2$,
 $h[S^{d_0}]$ is the canonical metric on $d_0$-dimensional sphere
 $S^{d_0}$,
  $H = 1 + P  R^{-d}$, $P(P + 2 \mu)= - 4 A/d^2 $ ($A < 0$, $P > 0$, $\mu >
 0$),  $d = d_0 -1$ and $r = 1/3, 2/3, 1/2$
  for $M2$, $M5$, $D3$  branes, respectively. In the
  extremal case  $\mu \to +0$  the near-horizon limit
  gives  an exact solution
  for the flat space $(M_2 = \R^{d_2},{h}^{[2]})$
  \begin{eqnarray}
  g=  P^{r} \biggl\{ h[S^{d_0}] + \frac{4}{(d -2)^2} h[AdS^{d_2 + 2}]
    \biggr\},
 \label{5.4.13} \\
  \rho = - A  P^{- 2 - r},  \qquad P^2 = - 4 A/d^2,
   \label{5.4.14}
 \end{eqnarray}
  where
  $h[AdS^{k + 2}] = du \otimes du +
  e^{2u}(- dy^0 \otimes dy^0 + \sum_{i =1}^k dy^i \otimes dy^i)$
  is the metric of (the part of) the anti-deSitter space
  $AdS^{k +2}$. Thus, we are led to a static configurations defined
   on (the parts of) the product  spaces
   $S^{d_0} \times AdS^{d_2 + 2}$, i.e.
   $S^{7} \times AdS^{4}$, $S^{4}\times AdS^{7}$
   and $S^{5} \times AdS^{5}$ for
    $M2$, $M5$ and $D3$  branes, respectively.
    It should be pointed out that the solutions with
    $AdS^{k}$ factor spaces may be of interest due
    to possible application in a context of
    $AdS/CFT$ approach \cite{Mald,GubKlPol,Witten}
    (or its modifications).

 \subsection{The Hawking temperature}

 The Hawking temperature of the black hole (\ref{12}) (see also
 (\ref{12a})) may be calculated using the relation from
 \cite{York}. It has the following form:

\beq{6.1}
 T_H=   \frac{d}{4 \pi (2 \mu)^{1/d}}
  \prod_{s = 1}^m H_{s0}^{- h_s}, \eeq

  where $H_{s0}$, $s =1,2$, are defined in
 (\ref{2.2.1a}).

For the dyonic solution from the previous subsection we get

  \beq{6.2}
 T_H =   \frac{1}{8 \pi  \mu}
   (H_{10}H_{20})^{- 1/2}, \eeq

 where $T_H$ is a function of fluid parameters $A_s < 0$,
 $s =1,2$.

\section{Conclusions}

Here we have presented a family of spherically symmetric solutions
with horizon in the model with multi-component anisotropic fluid
with the equations of state (\ref{1.7}) and the conditions
(\ref{2.1}) imposed.  The metric of any solution contains $(n -1)$
Ricci-flat ``internal'' space metrics.

As in  \cite{IMp1,IMp2,IMp3} the solutions are defined up to
solutions of non-linear differential  equations (equivalent to
Toda-like ones) with certain boundary conditions imposed. These
solutions may have a polynomial structure when the matrix $A$ from
 (\ref{2.1}) is coinciding with the Cartan matrix
 of some semi-simple finite-dimensional  Lie algebra.

For certain equations of state (with $p_i = \pm \rho$) the metric
of the solution  may coincide with the metric of intersecting
black branes (in a model with antisymmetric forms without
dilatons). Here we have considered an example of simulating of
$M2-M5$  black brane (dyonic) solution  in $D=11$ supergravity
with intersection rules corresponding to the Lie algebra $A_2$. We
have also outlined  the analogues of non-marginal $M2$, $M5$ and
$D3$  black brane solutions. In the extremal case $\mu \to +0$ the
near-horizon limits of all these solutions were found.

 An open problem is to generalize this formalism to the case when
 scalar fields are added into consideration.
 In a separate paper  we also plan to calculate the post-Newtonian
 parameters $\beta$ and $\gamma$ corresponding to the 4-dimensional
 section of the metric (for $d_0 =2$) and analyze the thermodynamic
 properties of the black-brane-like solutions in  the model with MCAF.

\begin{center}
 {\bf Acknowledgments}
 \end{center}

 This work was supported in part by
 grant NPK-MU (PFUR)
 and   Russian Foundation for Basic Research
 (Grant  Nr. 09-02-00677-a.).


 \end{document}